\documentclass[twocolumn,epsfig,graphics,showpacs,floatfix,mathbbm]{revtex4-1}

\usepackage{amsmath,amsfonts,amssymb,graphics,graphicx,epsfig,color,times,bbm}
\usepackage{amsthm}
\usepackage{braket}

\newcommand{\me}{\mathrm{e}}

\newcommand{\per}{\mathrm{per}}
\newcommand{\diag}{\mathrm{diag}}
\newcommand{\arccot}{\mathrm{arccot}}

\newcommand{\cc}{\mathbbm{C}}

\newcommand{\rr}{\mathbbm{R}}
\newcommand{\id}{\mathbbm{1}}

\newcommand{\ps}{\ensuremath{p_{\mathrm{s}}}}

\newcommand{\UU}{\mathcal U}
\newcommand{\ic}{,}
\newtheorem{theorem}{Observation}
\newcommand{\1}{\mathfrak{1}}
\newcommand{\0}{\mathfrak{0}}
\newcommand{\figspace}[1]{}
\newcommand{\rank}{\mathrm{rank}}
\newcommand{\vac}{\mathrm{vac}}
 
\begin{document}

\title{On photonic controlled phase gates}

\author{K.~Kieling$^1$, J.~O'Brien$^2$, and J.~Eisert$^{1,3}$}

\affiliation{$^1$Institute of Physics and Astronomy, University of Potsdam, 14476 Potsdam, Germany}
\affiliation{$^2$Centre for Quantum Photonics, H.~H.\ Wills Physics Laboratory and 
Department of Electrical and Electronic Engineering, University of Bristol, Bristol BS8 1UB, UK}
\affiliation{$^3$Institute for Advanced Study Berlin, 14193 Berlin, Germany}
\date\today

\begin{abstract}
As primitives for entanglement generation, controlled phase gates take a central role in quantum computing. Especially in ideas realizing instances of quantum computation in linear optical gate arrays a closer look can be rewarding. In such architectures, all effective non-linearities are induced by measurements: Hence the probability of success is a crucial parameter of such quantum gates. In this note, we discuss this question for controlled phase gates that implement an arbitrary phase with one and two control qubits. Within the class of post-selected gates in dual-rail encoding with vacuum ancillas we identify the optimal success probabilities. We construct networks that allow for an implementation by means of todays experimental capabilities in detail. The methods employed 
here appear specifically useful with the advent of integrated linear optical circuits, providing stable interferometers on monolithic structures.
\end{abstract}

\maketitle

\section{Introduction}

Linear optical architectures  offer
the potential for reliable realizations
of small-scale quantum computing~\cite{KLM01}
In the recent past, numerous 
proof-of-principle demonstrations have relied on the precise state
manipulation that is available using linear optical elements. The 
development of better and brighter sources with good mode 
quality as well as new types of detectors have opened up new 
perspectives~\cite{Further} in state preparation
and manipulation, for six or more photons. 
Specifically, integrated optical circuits allow for
state manipulation with little mode matching problems in interferometers~\cite{Integrated}. 

Naturally, significant efforts has been devoted towards realizing instances
of quantum gates. As primitives for such small-scale computing, two-partite quantum gates delivering a 
controlled phase-shift of $\varphi$ have already been experimentally
demonstrated (see Refs.~\cite{HT02,RLBW02,PPBS} for $\varphi=\pi$ and Ref.~\cite{LBA+09} 
for general phases). In this note, we focus on linear
optical implementations of phase gates with arbitrary phases. In particular,
we will ask when arbitrary phases can 
be realised in the first place, and---one of the main figures of merit in linear optical
applications---what the optimum probabilities of success are, as any non-linear map is necessarily
probabilistic.
A realisation of such gates seems interesting from the perspective of 
\begin{itemize}
\item[(i)] gaining an understanding of the probabilistic character of 
quantum gates as well as 
\item[(ii)] serving as a proof of principle
realisation of a kind of quantum gate that has several 
applications in linear optical quantum information processing.
\end{itemize}

As far as the first aspect is concerned, 
one may well expect that there
is a trade-off between the notorious problem of having a small
probability of success and the phase that is being realised in
the gate. In fact, the study in Ref.~\cite{Eisert04}
suggests exactly such a behavior: 
the presented upper bounds to the probability of success
increase from the 
minimum at $\ps(\varphi=\pi)=1/4$ to $\ps(0)=1$. To
investigate such a trade-off is interesting in its own right
and helps in building intuition concerning the probabilistic behavior
of linear optics. One may well develop the intuition that ``large phases are costly'' as far
as the probability of success and hence the overhead or repetition are concerned.

Further, concerning the second aspect, there are several applications 
for which such a trade-off is relevant. In linear optical architectures,
it may be a good idea to have a smaller phase, if one only
has higher success probabilities. The new measurement-based
quantum computational models~\cite{GE07a} for example offer this 
perspective: One does not have to have controlled $\pi$ phase gates
to prepare cluster states, but one would in principle also get
away with smaller phases. This may well (but does not have to be)
a significant advantage when preparing resources for
measurement-based quantum computing different from cluster states~\cite{RB01,GE07a}.

Of course, in standard gate-based quantum computing, 
one will typically encounter all kinds of controlled phase gates. For example,
in the quantum Fourier transform~\cite{nielsen}, 
one has to implement several controlled phase gates.
They can again be decomposed into other sets of universal 
gates (like CNOT or CZ and local unitaries). But,
in terms of resource requirements, it is obviously an advantage to directly
implement the relevant quantum gates with phases in the range 
$0<\varphi<\pi$.
There are also interesting trade-offs
between resource requirements and success probabilities in a number of
related contexts, like non-local gates in distributed quantum computation~\cite{EJPP00,CDKL01Berry07}. Refs.~\cite{CDKL01Berry07}, for example,
study distributed controlled phase-gates which would need less entanglement
and succeed with a higher probability.
In the field of linear optics gates, numerical results on direct implementation of arbitrary two-qubit gates
are known (see Ref.~\cite{dmitry} and references therein).

Instead of resorting to decompositions in the circuit model, one could gain
from implementing unitaries in a fashion ``natural'' to the respective architecture at hand.
In the case of linear optics it means to leave the computational sub-space given by the
encoding of the qubits for the sake of taking a ``shortcut'' through higher dimensions~\cite{LBA+09}.
Given that the fundamental information carriers are implemented using bosonic modes, this will
occur when mixing those modes in beam splitter networks and is an inherent feature of
genuine linear optics implementations, in contrast to decompositions into standard gate sets.

Here we will study post-selected gates, so not genuine ``event-ready'' quantum
gates, but---as is common in linear optical architectures at least to date---those 
where one measures the output modes and whether the gate
actually succeeded is determined only {\it a posteriori} by accepting only those
outcomes which lie in the computational dual-rail subspace of the Hilbert space of $n$ photons on $2n$ modes.
Incorporating less constraints, these gates concern only a smaller number of modes
and are still within reach of current experiments.
In principle non-demolition measurements of the output would be required
for an event-ready gate.

\section{Controlled phase gates}
\subsection{Single beam splitter}
In a post-selected phase gate on four modes in 
the standard dual-rail encoding, two of the modes
are merely involved as ``by-standers'', in that their amplitude is 
compensated in exactly the same fashion as in Refs.~\cite{HT02,RLBW02,PPBS}.
In this section, we will hence 
concentrate on two modes forming the ``core'' of the
scheme, giving rise to a two-qubit dual-rail phase gate on four physical modes. The core
itself may be regarded as a single-rail phase gate in its own right. Later we will see that
not breaking the network into a core and by-stander modes will not give any advantage.

Similar to Ref.~\cite{HT02} we will briefly investigate the consequences of simply having a single beam
splitter forming the core of the quantum gate.
The action on the photonic creation operators of the two involved
modes it is mixing is described by the matrix
\begin{equation}
  U = \diag (\me^{\imath\phi_1},\me^{-\imath\phi_1}) \cdot B \cdot \diag (\me^{\imath\phi_2},\me^{-\imath\phi_2} )
\end{equation}
with
\begin{equation} \label{eqn:orthogonal}
  B = \left[\begin{array}{cc} \sin(\vartheta) & \cos(\vartheta) \\ -\cos(\vartheta) & \sin(\vartheta) \end{array}\right] 
\end{equation}
and appropriate phases $\phi_1,\phi_2\in[0,2\pi)$
and mixing angle $\vartheta\in [0,2\pi)$.
The phases can also be realised deterministically by local operations on the dual-rail qubits, which leaves
the relevant part of the gate $U'=B$. The matrix elements of the unitary $\UU'$ belonging to $U'$ for vacuum, single photon operation, and the two-photon component
read
\begin{eqnarray}
  \braket{0,0|\UU'|0,0} &=& 1 ,\label{eqn:cphase:def1} \\
  \braket{1,0|\UU'|1,0} &=& A_{1\ic 1} = \sin(\vartheta), \\
  \braket{0,1|\UU'|0,1} &=& A_{2\ic2} = \sin(\vartheta), \\
  \braket{1,1|\UU'|1,1} &=& \per( A) = \cos(2\vartheta) = 1-2\sin^2\vartheta, \label{eqn:cphase:def2}
\end{eqnarray}
respectively.
Since we are restricted to $n\le2$ modes, these four quantities determine the action of the core completely.
With the constraint
\begin{equation}
  1-2\sin^2(\vartheta) = \sin^2(\vartheta)
\end{equation}
that ensures equal single- and two-photon amplitudes (equal probabilities for all dual-rail states),
only the two solutions $\vartheta=\pm\arcsin(3^{-1/2})$ are possible, giving rise to
$\varphi=0,\pi$, respectively.

Hence, one finds that in this way, one \emph{can} implement quantum phase gates, but only
two different ones: One is not doing anything, and the other ones effect is a controlled
phase of $\pi$. This is exactly the gate of Refs.~\cite{HT02,RLBW02}. In other words, 
without invoking at least a single additional mode, one can not go beyond
the known $\pi$-phase in this fashion.

\subsection{Arbitrary phases}
However, we can extend this scheme: 
The restriction to unitary two-mode beam splitters can be relaxed.
Instead of starting with $U\in SU(2)$, we use an arbitrary matrix $A\in\cc^{2\times2}$.
Then we will embed the two-mode matrix into a higher dimensional unitary, such that an 
appropriately rescaled $A$ forms a principle submatrix of a larger unitary matrix $A'$. The optimal rescaling is simply
dictated by the largest singular value of $A$. We will see that in the two-mode case only a single additional mode is already the most general extension, so the
full set-up would consist of a transformation on three modes involving 
at most three beam splitters. In this class of gates, for each $\varphi$, 
the one with the optimal probability of success $\ps(\varphi)$
can be identified.

\begin{figure}
  \includegraphics{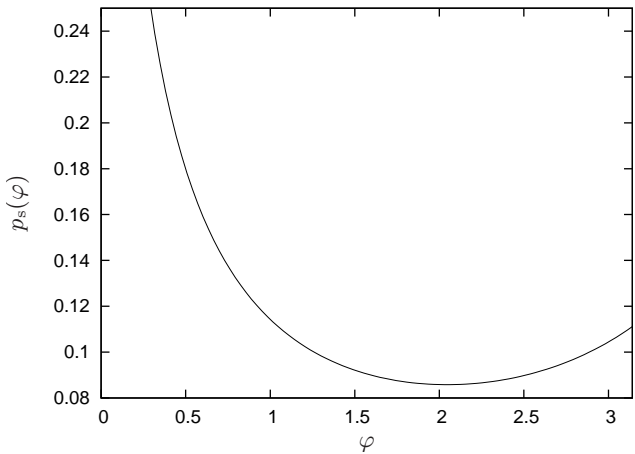}
  \caption{Optimal success probability $\ps(\varphi)$ of phase gates with vacuum ancillas (one vacuum ancilla is already optimal)
    {\it vs.}\ the phase $\varphi$ (solid line). At $\varphi=\pi$ the result of Refs.~\cite{HT02,RLBW02},
    $\ps(\pi) = 1/9$, is reproduced.
    The intuitive assumption of a monotonous $\ps(\varphi)$ is not fulfilled: indeed, the success probability is worse than $1/9$ in the interval $\pi/3 < \varphi < \pi$.
    Due to phases $\varphi<\pi$ not being implementable with a single beam splitter, the additional unitary extension requires further measurements and therefore decreases the probability
    of success near $\varphi=\pi$.
    \label{fig:phases} }
\end{figure}

\begin{theorem}[Optimal post-selected dual-rail controlled phase gate]
  Consider linear optics, an arbitrary number of auxiliary vacuum modes and photon number resolving detectors. When post-selecting the state of the signal modes onto the computational sub-space
  and the auxiliary modes onto the vacuum,
  the optimal network on four modes implementing the gate represented in the computational basis of two dual-rail qubits by
  $U = \diag ( 1,1,1,\me^{\imath\varphi} )$, $\varphi\in[0,\pi]$,
  has a success probability (shown in Fig.~\ref{fig:phases}) of
  \begin{equation}
    \ps(\varphi) = \left(1+2\left|\sin\frac{\varphi}{2}\right|+2^{3/2}\sin\frac{\pi-\varphi}{4}\sqrt{\left|\sin\frac{\varphi}{2}\right|}\right)^{-2} .
  \end{equation}
\end{theorem}

{\it Proof:}
In order to find $\ps$---the same for all possible input states---we will first construct the linear transformation of the relevant creation 
operators and identify the optimal unitary extension afterwards.
\begin{itemize}
\item 
{\it Two-mode transformation:}
The two-mode transformation resulting from solving the equations (\ref{eqn:cphase:def1})--(\ref{eqn:cphase:def2}) imposed by the gate we want to build is
\begin{equation}\label{eqn:cphase:solution}
  A = \ps^{1/4}\left[ \begin{array}{cc} x & \left(\me^{\imath\varphi} - 1\right)x/y \\ y/x & 1/x  \end{array} \right] .
\end{equation}
$x$ and $y$ are free non-zero complex parameters. By writing
\begin{equation}
  A = \ps^{1/4}\,\diag (a,a^{-1})
  \cdot\left[ \begin{array}{cc} 1&\me^{\imath\varphi}-1 \\ 1&1 \end{array}\right]\cdot\diag(b,b^{-1}) ,
\end{equation}
with $a=xy^{-1/2}$ and $b=y^{1/2}$ we see that the singular values of $A$ only depend on $|a|^2$ and $|b|^2$, so not on the phases of $a$,$b$, and $x$ and $y$.

The general solution to the dual-rail problem is actually composed of the transformation $A$ together with appropriate damping of the by-standers:
the probability of success can not be enhanced by considering a full transformation on all four modes.
This can be seen by writing the polynomial system given by the dual-rail problem similar
to (\ref{eqn:cphase:def1})--(\ref{eqn:cphase:def2}), consisting of $16$ quadratic equations
in the matrix elements of $B\in\cc^{4\times 4}$. It turns out that by permuting modes and appropriate variable substitutions all solutions can be brought into the form
$B \propto \id_2 \otimes A$,
consisting of the two-mode core given in Eqn.~(\ref{eqn:cphase:solution}) and two by-passed modes.

\item {\it Optimal extension:}
Given the $2\times2$ matrix $A$ that realises the transformation we are looking for,
the optimal unitary 
extensions can be identified. Let us extend the first and second row vectors (denoted by $A_1$ and $A_2$) to dimension $3$ by appending $A_{1\ic3}$ and $A_{2\ic3}$,
respectively, in such a way as to allow for unitarity of the extended matrix, $A'\in SU(3)$.


To see why a dimension of three is already sufficient consider a linear transformation of the creation operators of $n$ modes, described by its a (not necessarily unitary) matrix $A$.
By using the singular value decomposition (SVD)
it can be decomposed as $A=V\cdot D\cdot W^{-1}$ where $V$ and $W$ are unitary (and therefore have immediate interpretations as physical beam splitter matrices themselves),
and $D=\{d_1,\ldots,d_n\}$ is a diagonal matrix with real non-negative entries $d_1\ge d_2\ge\ldots\ge d_n$, the singular values.
In terms of linear optics $D$ can be interpreted as mixing each mode $k=1,\ldots,n$ with an additional
mode $n+k$ in the vacuum state which will be post-selected in the vacuum afterwards~\cite{MLU+06,Kieling08}.
Then, $d_k$ describes the transmittivity of the beam splitter used to couple modes $k$ and $n_k$.
Without loss of generality one can assume $d_1=1$, vacuum mixing for the first mode.
This can be achieved by rescaling with the inverse of the largest singular value, so $A \mapsto d_1^{-1}A$, which implies $d_k\mapsto d_1^{-1}d_k$.
Note that such a ``global'' rescaling of $A$ does not change the post-selected action on the computational sub-space but only the success probability of it according to $\ps\mapsto d_1^{-2n}\ps$.
Therefore, in general, there are $n-1$ additional
vacuum modes required to extend an $n$-mode linear transformation to a unitary, and thus physical, network. Also please note, that the constraint $d_1=1$
has to be taken into account for any optimisation of success probabilities of $A$. Here we will not explicitly use this decomposition further, but the constraint will be implemented
implicitly by requiring the $n-1$-mode extension to be unitary.


In our specific $3$-mode extension we choose
$A_{1\ic3}$ and $A_{2\ic3}$ such that the new row vectors are orthogonal. By multiplying them by the root of the inverse of their respective norms, $|A'_1|$ and $|A'_2|$,
they will be normalised. Finding a third orthogonal vector to fill the unitary matrix can be done with the complex cross product $(A'_1\times A'_2)^*$, or
in general by choosing a vector at random and orthogonalising it with respect to the given ones.

The dependence of the success probability on the extension is 
\begin{equation}
	\ps=\left(|A'_1||A'_2|\right)^{-2}.
\end{equation}	
Therefore, the objective is to
\begin{eqnarray}
  \mbox{minimise}   && f = |A'_1|^2 |A'_2|^2 \\
  \mbox{subject to} && A'_1 (A'_2)^{\dagger} = A_1 A_2^{\dagger} + A_{1\ic3}A^{*}_{2\ic3} = 0 \label{eqn:subject} .
\end{eqnarray}
The first observation is that the row-scaling by $x$ is already included in the norm of the row vectors, leaving us with one parameter less.
By using the phase of $y$, we can assure that $A_1 A_2^{\dagger}$ is real and 
positive and also $\arg(A_{1\ic3}) - \arg(A_{2\ic3}) = \pm\pi$. This 
constrained minimisation problem in $A_{1\ic2}$ and $A_{1\ic3}$ can indeed be solved
(by using Lagrange's multiplier rule and showing constraint qualification)
and  we find $|y|=\left(2(1-\cos\varphi)\right)^{1/4}$.
Then an optimal solution (phases chosen conveniently) is
\begin{equation}
  A_{1\ic3} = A^*_{2\ic3} = \me^{{\imath\pi}/{2}}\left(\sqrt{2\left|\sin\frac{\varphi}{2}\right|}\sin
  \frac{\pi-\varphi}{4}\right)^{1/2}
\end{equation}
with the probability of success given by
\begin{eqnarray}\label{eqn:cphase:ps}
  \ps(\varphi) &=& \left(1+|y|^2+|y|\sqrt{2-|y|^2}\right)^{-2} \\
      &=& \left(1+2\left|\sin\frac{\varphi}{2}\right|+2^{3/2}\sin\frac{\pi-\varphi}{4}\sqrt{\left|\sin\frac{\varphi}{2}\right|}\right)^{-2} .
      \nonumber
\end{eqnarray}
The reflectivities of the compensating beam splitters in the by-passed modes have to be chosen such that the success probability is constant for all dual-rail states, i.e., 
\begin{equation}
	r=\ps^{1/4}.
\end{equation}	
\end{itemize}

\begin{figure}
  \includegraphics{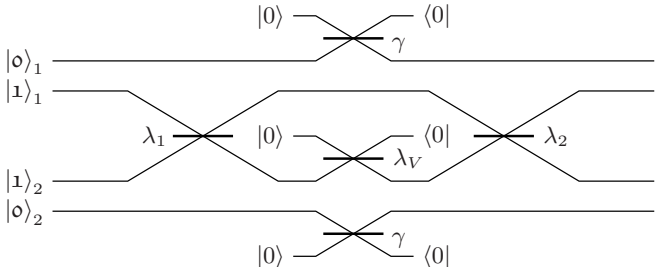}
  \caption{Basic spatial modes based set-up obtained from translating an arbitrary $2\times 2$ core into the language of linear optics. 
   The core extension is provided by mixing
    with a vacuum mode on the central beam splitter. This mode, in turn, has to be post-selected in the vacuum state afterwards. The upper and lower beam splitters
    implement the appropriate compensation by damping the by-passed modes (which is the same for both modes for the optimal solution which we consider).
\\
    The labels at the beam splitters will be used to identify them with the respective optical elements in Figs.~\ref{fig:pgate03}--\ref{fig:pgate:ppbs}.
    In general, the parameters ({i.e.}, reflectivity and phases) of these elements depend on the gate's phase $\varphi$.
    Further, the notation $\ket{\0}_i$ and $\ket{\1}_i$ is used for the logical $0/1$-modes of the $i$-th qubit to avoid confusion with Fock states. \label{fig:pgate01} }
\end{figure}

The success probability $\ps(\varphi)$ of this gate
is shown in Fig.~\ref{fig:phases}. 
Interestingly---and quite surprisingly---the 
worst success probability is not achieved for the sign-flip ($\varphi=\pi$), but for
$\varphi\approx2.05$. This means, gates delivering a phase shift slightly
smaller than $\pi$ and thereby generating less entanglement will not give rise to a larger, but to a smaller success probability.
As expected, the success probability for very small 
phases increases and reaches unity for $\varphi=0$: one can always do nothing 
at all with unit success probability.


\subsection{Integrated quantum photonics realizations}
\begin{figure}
  \figspace{\vspace{.25cm}}\includegraphics{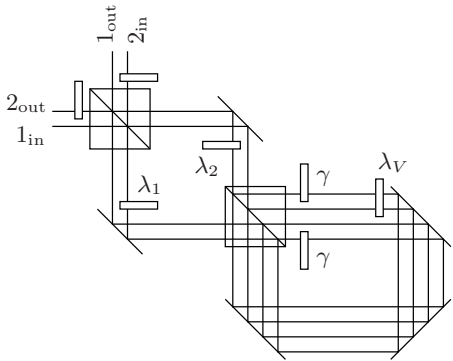}
  \caption{Compact implementation of a controlled phase gate by using a single loop to implement the central PPBS and the compensation beam splitters simultaneously. Additionally, the two PBS are identified,
     resulting in a second loop. All omitted modes are initialised in the vacuum and post-selected in the vacuum state (which will be achieved in practice by counting the photons in the other output).
     \label{fig:pgate:ppbs} }
\end{figure}

Sophisticated circuits such as the one shown in Fig.~\ref{fig:pgate:ppbs} can be built from 
bulk optical elements (mirrors, beamsplitters, etc.) Such circuits often
require implementation of Sagnac interferometers (e.g., Ref.~\cite{NOO+07}), partially polarizing beam splitters (PPBSs)~\cite{PPBS}, or
beam displacers~\cite{OPW+03,Process1} to achieve interferometric stability. For the most complicated circuits a combination of these elements is
required. Indeed a (non-optimal) implementation of a two-qubit controlled unitary gate used a combination of beam displacers and
PPBSs~\cite{LBA+09}. However, such circuits are extremely challenging to align, are limited in performance by the quality of that alignment, and are ultimately not scalable.

An alternative approach based on lithographically fabricated integrated waveguides on a chip has recently been developed~\cite{Integrated}.
This approach has demonstrated better performance, in terms of alignment and stability, as well as miniaturization and scalability.
The monolithic nature of these devices enables interferometers to be fabricated with precise phase and stability, making the Mach-Zehnder
interferometer shown in Fig.~\ref{fig:pgate01} directly implementable without the need to stabilize the optical phase (either actively,
or using the Sagnac-type architecture of Fig.~\ref{fig:pgate:ppbs}), greatly simplifying the task of making complicated circuits:
Essentially the circuit one draws on the blackboard can be directly `written' into the circuit.
Indeed, integrated photonics circuits have been used to implement a circuit of several logic gates on four photons, to implement a compiled
version of Shor's quantum factoring algorithm in this way~\cite{PMO09}. In fact laser direct write techniques have been
used to `write' circuits in an even more direct fashion than the lithographic approach.

Another key advantage of integrated quantum photonics for the circuits described here is that on-chip phase control can be directly
integrated with the circuit~\cite{IntegratedReconfigurable}, which could allow measurment of the success probability
curve (Fig.~\ref{fig:phases} or Fig.~\ref{fig:toffoli:ps} for example)
to be directly mapped by sweeping the applied voltage.

\subsection{Experimental issues in free space}
\begin{figure}
  \figspace{\vspace{.25cm}}
  \includegraphics{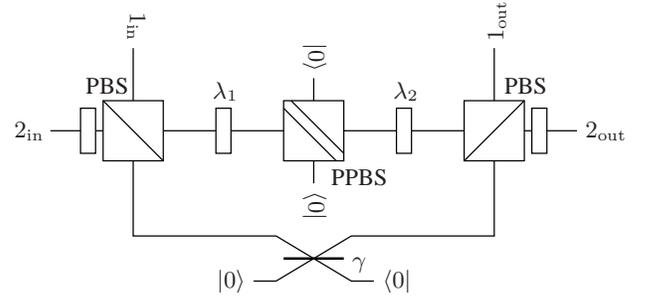}\figspace{\hspace{.25cm}\vspace{.5em}}
  \caption{Setup for a controlled phase gate on two polarisation encoded dual-rail qubits. The logical modes of the two qubits are separated and re-united by means of polarising beam splitters (PBS).
    Replacing the left and right beam splitters in Fig.~\ref{fig:pgate01} by wave plates $\lambda_1$ and $\lambda_2$,
    they become easier to tune to different $\varphi$, and provide better stability. The lower beam splitter, $\gamma$,
    implements compensation of both, $\ket{\0}_1$ and $\ket{\0}_2$, modes. 
    $\lambda_V$ is taken care of by the partially polarising beam splitter (PPBS, allowing for 
    different reflectivities for the two polarisations, further explained in Fig.~\ref{fig:ppbs}) in the centre.
    Additionally, one of the qubits has to be flipped prior to and after the circuit, which here is done by acting with a wave plate on the second qubit. \label{fig:pgate03} }
\end{figure}

In order to render the proposed gates experimentally more feasible in free space, some simplifications
have to be done -- tailored to the specific physical implementation at hand.
Waveguide based setups would not need further simplifications because stablity of the interferometers would be ensured by the rigid substrate.
Implementations more suitable for beam displacer based setups are known~\cite{LBA+09}.
Influenced by the gate model, those gates include a controlled $\pi$-phase gate at the core. Operating at lower success probabilities,
especially at phases $\varphi$ approaching $0$, the probability of success does not converge to $1$.

In the following we will discuss simplifications which shall allow for easier free-space implementation of the networks introduced above
while still preserving optimality with respect to $\ps$.
\begin{figure}
  \includegraphics{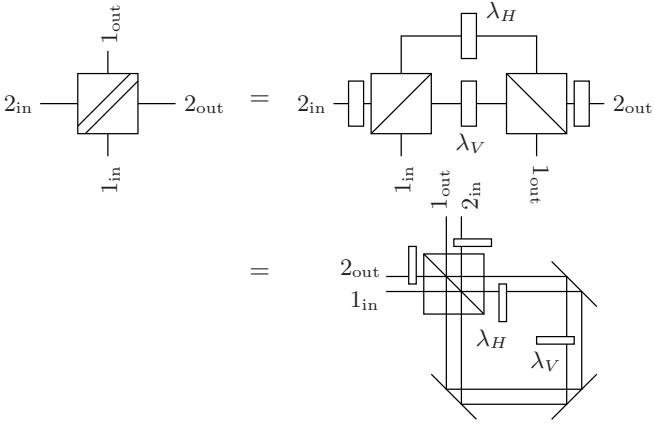}
  \caption{From left to right: (i) A PPBS implementing a beam splitter with polarisation dependent reflectivity. (ii) It is equivalent to an interferometer between two PBS
    where the PPBS' reflectivities are incorporated by means of wave plates, $\lambda_1$ and $\lambda_2$. (iii) By identifying the two PBS, the interferometer collapses into a closed
    loop (which is more compact and more robust in experimental implementations), leaving only one PBS.
    In the second and third circuit, a polarisation flip of the second qubit before and after the circuit is added by using a wave plate.\label{fig:ppbs} }
\end{figure}
A straightforward set-up on dual-rail encoding that realises a three-mode unitary and 
compensates the amplitudes in the remaining modes is shown in Fig.~\ref{fig:pgate01}.
It includes one interferometer, but the whole gate would sit inside a double interferometer,
because local unitaries on the input and output qubits would require classical interference.
Thus, the complexity of this gate is best described as a nested three-fold interferometer.
In this first stage, the parameters (reflectivities and phases) of all five beam splitters
depend on $\varphi$.

To get rid of some of the interferometers, polarisation encoding is convenient. Two modes can be united in one spatial mode, resulting in inherent stability (neglecting
birefringence of the optical medium) of some interferometers. Rotations on these modes can be carried out easily using wave plates. Because the core acts on modes coming from different
dual-rail qubits, they have to be combined into a single spatial mode before. This is achieved with a PBS, thus permuting $H$ and $V$ modes. Damping of both by-passed modes can be done simultaneously
by a single polarisation insensitive beam splitter coupled to the vacuum. A straightforward translation of Fig.~\ref{fig:pgate01}
into polarisation encoding, thereby collapsing the network into a single interferometer, is shown in Fig.~\ref{fig:pgate03}.

Due to the asymmetry of the core, still one PPBS is used, the reflectivity of one polarisation component of which actually depends on $\varphi$, the other one being $1$. Fig.~\ref{fig:ppbs} shows
how a tunable PPBS can be constructed, introducing another interferometer.

Iterating the ideas that led to the compact PPBS implementation once more yields a collapsed form of the phase gate based on only two PBS and a couple of wave plates. Due
to the paths of light being very similar, this set-up should also be more robust. 
This is illustrated by Fig.~\ref{fig:pgate:ppbs} since it has an overall Mach-Zehnder interferometer structure to it, by feeding one displaced sagnac loop into another.

\subsection{Simple decomposition}
Here we want to obtain a simple decomposition and explicit $\varphi$-dependence of the involved elements of the full $3\times3$ unitary transformation $U\in SU(3)$.
To do so, we interpret the effective core, acting on the $\ket{\1}_1$ and $\ket{\1}_2$ modes in Fig.~\ref{fig:pgate01}, in a different way:
in between $U_{\lambda_1}$ (the unitary beam splitter matrix of $\lambda_1$)
and $U_{\lambda_2}$, there acts a diagonal mode transformation such that the first mode is unaffected and the amplitude of the second one is damped
(due to $U_{\lambda_V}$ coupling it with reflectivity $r_V$ to a vacuum mode which will be projected onto the vacuum).

Now considering the singular value decomposition (SVD) of the matrix representing the core transformation, $A=V\cdot\Sigma\cdot W^{\dagger}$ we can identify $V=U_{\lambda_2}$,
$W^{\dagger}=U_{\lambda_1}$ and $\Sigma=\diag\{1,r_V\}$. As we have seen earlier, optimal extensions of $2\times2$ cores only require global rescaling, which commutes with the unitaries involved.
Therefore we can use the much simpler original form of $A$ in Eqn.~(\ref{eqn:cphase:solution}), and we choose $x=1$ and $y=-\sqrt{\me^{\imath\varphi}-1}$.

We find the singular values of $A$
\begin{equation}
  \sigma_{\pm} = \sqrt{1+2\sin\frac{\varphi}{2}\pm2(2-2\cos\varphi)^{1/4}\cos\frac{\varphi+\pi}{4}} .
\end{equation}
Global rescaling amounts to fixing the largest singular value to unity, so the new singular values are $1$ and $r_V=\sigma_-/\sigma_+$.

Due to $\det U_{\lambda_1}=\det U_{\lambda_2}=1$
we need to attach a phase to one singular value as well in order to apply the identification of the matrices introduced above. 
Then the SVD of $A$ yields
\begin{equation}
  U_{\lambda_1} = \frac{1}{\sqrt{2}}\left[ \begin{array}{cc} -1 & 1 \\ -1 & -1  \end{array} \right]
\end{equation}
and $U_{\lambda_2} = U_{\lambda_1}^{-1}\cdot\me^{\imath\phi_+\sigma_z}$ with
\begin{equation}
  \phi_\pm=\arccot\left[ \cot\frac{\varphi+\pi}{4}\pm\left((2-2\cos\varphi)^{1/4}\sin\frac{\varphi+\pi}{4}\right)^{-1}\right]
\end{equation}
where the order of rows and columns in the matrices is as in Fig.~\ref{fig:pgate01} from top to bottom.

The ``complex singular values'' are $1$ (the first mode is not affected) and $\sigma_-/\sigma_+\exp{\imath \phi_++\phi_-}$. The latter can be achieved by using the aforementioned
coupling to the vacuum with a reflectivity of $r_V$ and a phase upon reflection of $\phi_++\phi_-$. The further ingredients are the beam splitters required for the ``damping'' of the
by-standers as discussed earlier.
By confirming $1/\sigma_+^4 = \ps$ in the range $0\le\varphi\le\pi$, the optimality of this construction is assured.

\section{Event-ready gates}
Coming from post-selected gates, the next step towards scalable quantum computation would be to build
gates not requiring measurements on the output modes. Intuitively it is clear that the construction of
a controlled phase gate in this class will be more demanding with respect to the resources (such as
the number of auxiliary modes and photons, size and complexity of the network) involved.

Especially the number of additional photons will change drastically: having had none in the post-selected case of a controlled $\pi$-phase gate,
two are required in the class of event-ready gates. We will use this example as a motivation for a detour to discussing a number
of different methods that could be useful for handling linear optics state preparation.

To do so, we notice that a controlled $\pi$-phase gate is more constrained than
a device that creates EPR pairs from single photons.
This is meant in the sense that it not only amounts to
a state transformation from two single photons to an EPR pair, but a full unitary transformation
on the entire computational state space in dual-rail encoding (and creating an EPR pair when applied to a
certain product input corresponding to the product state of two photons).
In the following two sections we will be concerned with different methods to describe linear optics state preparation
and will apply them to the specific example at hand (i.e., heralded dual-rail EPR pair generations from single photons).

It will turn out, that the construction of an EPR pair out of single photons by means of linear optics, vacuum modes,
one additional photon, and detectors is not possible.
Of course, directly solving the polynomial equations in the matrix elements of $A$ (generalisation of Eqns.~(\ref{eqn:cphase:def1}) to~(\ref{eqn:cphase:def2})) will yield the same result -- no solutions unless two
additional photons are involved.
Having excluded the cases of zero and one auxiliary photons, a set-up with two of them 
is possible, proven by the existence of such a scheme (EPR construction~\cite{ZBL+06} as well
as controlled-$Z$ gate~\cite{KLM01}).

\subsection{State transformations}
An obvious way of looking at states of exactly $2$ photons in $m$ bosonic field modes is the following. 
Such a state vector can be written as
\begin{eqnarray}
  |\psi_M\rangle &=& P({\bf a}^\dagger)|\vac\rangle \nonumber\\ &=& \sum\limits_{i,j=1}^{m} M_{i,j} 
  a_i^\dagger a_j^\dagger |\vac\rangle = ({\bf a}^\dagger )^T M {\bf a}^\dagger |\vac\rangle ,
\end{eqnarray}
where $M$ is a symmetric $m\times m$ matrix~\footnote{The $m=4$ case was used already in 
Refs.~\cite{LCS99,Calsamiglia02}. Thanks to D.~Uskov for pointing out the nice structure of this.}.
The application of a unitary mode-transformation $U$---representing a linear optical network---is reflected by
\begin{eqnarray}
  |\psi_M\rangle \mapsto |\psi_{M'}\rangle &=& (U{\bf a}^\dagger)^T M (U{\bf a}^\dagger) |\vac\rangle \\
                  &=& ({\bf a}^\dagger)^T M' {\bf a}^\dagger |\vac\rangle
\end{eqnarray}
with $M'=U^T MU$ clearly again being symmetric.
As a special case of the singular value decomposition~\cite{HornJohnson}, a diagonal $M'$ can be achieved,
given an arbitrary input state vector $|\psi_M\rangle$.

Now let us choose $U$ such that $M'$ is diagonal. Then, labelled by
\begin{equation}
  \nu' = \rank (M'), 
\end{equation}	
there are $m$ different classes of states~\cite{Kieling08,Kieling09}, in each of which is the states are composed by superpositions of $2$ photons in either of $\nu'$ modes.
These classes are separated by linear optical mode transformations requiring additional modes.
Decreasing the rank is possible by allowing for auxiliary vacuum modes. However, to increase the rank by $1$ one additional photon is required.

Further, the state matrix $M$ of two single photons on four modes has rank $\nu=2$ while an EPR pair corresponds to a matrix with rank $\nu=4$.
Therefore, the desired state transformation requires at least two additional single photons.

\subsection{Polynomial factorisation}
An alternative approach is the following~\cite{Kieling08}: The polynomial describing 
the objective state vector $|\psi\rangle = P({\bf a}^\dagger)|\vac\rangle$
with
\begin{equation}
  P({\bf a}^\dagger) = 2^{-1/2}\left(a^{\dagger}_1 a^{\dagger}_3 + a^{\dagger}_2 a^{\dagger}_4\right)
\end{equation}	
does not factorise over $\cc$.
Using Lemmata~1 and~2 from Ref.~\cite{Ruppert99}, the property of factorisation of a bivariate polynomial 
\begin{equation}
  p(x,y)=\sum_{i,j=0}^mp_{i,j}x^iy^j 
\end{equation}
over $\cc$ can be tested by assessing the rank of a 
complex $2m(2m-1)\times(m+1)(2m-1)$ matrix. Further, 
applying Lemma~7 of Ref.~\cite{Kaltofen95}, this technique can be extended to multi-variate polynomials.
Now, a state can be constructed from a product state using linear optical gate arrays 
iff the corresponding polynomial is factorisable.
In the case mentioned before (dual-rail EPR pair, so four variables), one can use the resulting $12\times 9$ matrix to confirm in the language of polynomials of creation operators that
additional resources are in fact required.

\begin{figure}
 \includegraphics{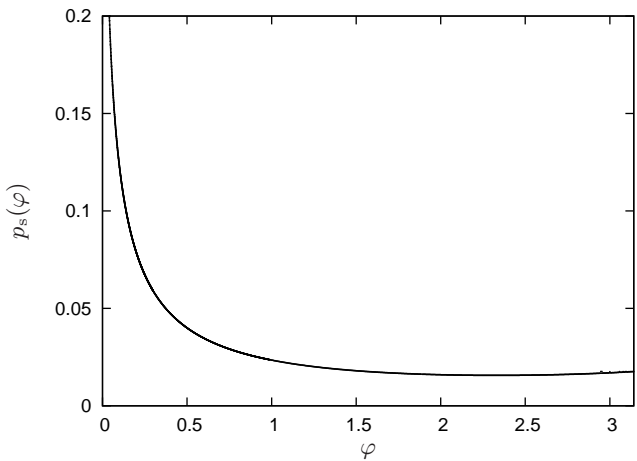}
  \caption{Optimal success probabilities of generalised Toffoli gates.
    The features exhibited by $\ps(\varphi)$ are similar to the ones observed at the controlled phase gate (Fig.~\ref{fig:phases}):
    There is a shallow dip between $\varphi=\pi/2$ and $\varphi=\pi$ below $\ps(\pi)=\ps(\pi/2)$ and a steep incline (more pronounced than for the controlled phase gate) for small phases towards $\ps(0)=1$.
    \label{fig:toffoli:ps}}
\end{figure}

\section{Toffoli gates}
In the same way as above, we can consider a generalised Toffoli gate, the effect of which on the 
computational basis realized as dual-rail encodings can be described by the unitary 
\begin{equation}
	U=\diag(1,1,1,1,1,1,1,\me^{\imath \varphi}).
\end{equation}	
The solutions to the polynomial equations describing the action on the three-mode core---up to mode permutations---can be parametrised by $x,y\in\rr$ and are given by the matrix
\begin{equation}
  A=\ps^{1/6} \left[\begin{array}{ccc} 1 & \frac{\me^{\imath\varphi}-1}{xy} & 0 \\ 0 & 1 & y \\ x & 0 & 1 \end{array}\right] .
\end{equation}
This has to be understood similar to the $2$-mode core used by the controlled phase gate. However, here we do not solve the full optimisation problem, but only consider global rescaling.

For a unitary extension all singular values of $A$ have to be at most $1$. In order to avoid to formulate 
cubic singular values explicitly, we use the following constraints. Let
$p_{AA^{\dagger}}(\lambda)=\det(A A^{\dagger}-\lambda\id_3)$ be the characteristic polynomial of $A A^{\dagger}$, the roots $\lambda_{1,2,3}$ of which are the squared singular values of $A$.
By requiring $p_{AA^{\dagger}}(1)=0$, one of the singular values has to be $1$. For the roots of $p_{AA^{\dagger}}$ are real-valued and non-negative,
the condition that all other singular values are not larger 
than $1$ is equivalent with the condition
that all derivatives of $p_{AA^{\dagger}}$ have the same sign at $\lambda=1$. 
More formally, this results in further constraints of the form
\begin{equation}
  (-1)^{\alpha} p^{(k)}_{AA^{\dagger}}(1) = (-1)^{\alpha} \left.\frac{\mathrm{d}p_{AA^{\dagger}}(\lambda)}{\mathrm{d}\lambda}\right|_{\lambda=1} \ge0
\end{equation}
for $1\le k<n$ where $\alpha=0,1$ is fixed by the condition of $k=n$.
For $\varphi=\pi$ the optimal $\ps$ compliant with these conditions is
\begin{equation}
  \ps(\pi) = 1 + 3\left( 2^{1/3}-2^{2/3} \right) \approx 1/57 .
\end{equation}
See Fig.~\ref{fig:toffoli:ps} for the maximum success probability in the range $0\le\varphi\le\pi$.
The corresponding networks could be constructed in the same way as above. However, they would consist of $3$-mode cores
(an interferometer composed of three partially polarising beam splitters)
inside separate interferometers
for each of the $3$ qubits.
For free space experiments more appealing approaches for the specific choice of 
$\varphi=\pi$ are presented in Refs.~\cite{RRG07,Fiurasek06}, but 
leading only to success probabilities of at most $1/72$.

\section{Remarks on process tomography}

Process tomography amounts to characterizing (or sometimes certifying) 
an unknown physical process. In practice, the task is to identify that completely
positive map that is closest to the date with respect to some meaningful figure of merit.
To accomplish this task, 
one has to consider a tomographically complete set of inputs and look at Hilbert-Schmidt
scalar products of the output with observables, to faithfully reconstruct the matrix form of the channel 
\cite{Process1,Process3}. 
Practically, the closest physical process can then be found by solving
a convex optimization problem. 
Formally equivalently, and in instances in an experimentally simpler fashion, one can, instead of
sending in a full set of input states, submit half of a single fixed maximally entangled state, and hence 
reconstruct the channel from the Choi matrix, then referrred to as entanglement-assisted 
process tomography.

The latter technique can clearly also be applied in case of a postselected quantum gate
like a phase gate. Yet, even without entangled inputs and including the actual measurement,
one can  reconstruct the resulting POVM elements, to which essentially postselected gates
amount to when one faithfully includes also the actual measurement in the black-box description
of the process. If one has a well-characterized source at hand, then the statistics of
\begin{equation}
	p_{j,k}= \text{tr}[\rho_j A_k],
\end{equation}
uniquely characterize the process,
where $\{\rho_j\}$ form a tomographically complete well-characterized
input set, and $A_1,\dots, A_K\geq $ constitute a POVM, i.e., 
\begin{equation}
	\sum_{k=1}^K A_k=\id
\end{equation}
(for an experimental realization of such an approach, see Ref.\ \cite{Process2}). In practice,
one uses methods of convex optimization to identify
the closest physical process to the given data.

For the purposes of the present work, 
one of the outcomes $k=1$, that is, a specific pattern $A_1$
of detection, then gives rise to the actual postselected linear
optical quantum gate. In this way, one can reconstruct
postselected quantum gates, without using an ancilla-based approach,
even faithfully including the final measurement as part of the process.

\section{Conclusion}

We have shown how to obtain the maximum probability of success of controlled phase- and Toffoli-like gates in 
the class of post-selected linear optics dual-rail gates without additional photons.
Further, constructions of networks for the smaller gates suitable for experimental implementation have been 
given, and techniques elaborated upon that allow for the assessment of the possibility
of certain linear optical schemes.
For further progress concerning the eventual full optimization of linear optical processes 
it would be interesting
to investigate (i) optimal constructions with respect to a class of gates inherently incorporating such 
experimental constraints or (ii) identify further
decomposition techniques from a given linear optics mode transformation
into suitable physical networks, respecting these constraints.

\section*{Acknowledgements}

For insightful discussions concerning current 
experimental 
abilities we would like to thank
M.~Du{\v{s}}ek,
W.~Wieczorek,
C.~Schmid, and
J.~Matthews. 
KK was supported by Microsoft Research through the 
European PhD Programme and the EU (MINOS), 
JE by the EU (QAP, COMPAS, MINOS) 
and the EURYI award.

\end{document}